\documentclass{article}
\usepackage{pgfplots}
\usepackage{spconf,amsmath,graphicx}
\usepackage{booktabs}
\usepackage[english]{babel}
\usepackage[utf8]{inputenc}
\usepackage{subcaption}

\usepackage{algorithm}
\usepackage[noend]{algpseudocode}
\algrenewcommand\algorithmicrequire{\textbf{Input:}}
\algrenewcommand\algorithmicensure{\textbf{Output:}}

\title{Language-Based Audio Retrieval with Converging Tied Layers and Contrastive Loss}

\name{Andrew Koh, Chng Eng Siong}
\address{Nanyang Technological University\thanks{This research is supported by ST Engineering Mission Software \& Services Pte. Ltd under collaboration programme (Research Collaboration No: REQ0149132).}}
\pgfplotsset{compat=1.17}
\tikzstyle{line}=[draw] 
\begin{document}
\maketitle
\begin{abstract}
In this paper, we tackle the new Language-Based Audio Retrieval task proposed in DCASE 2022\footnote{https://dcase.community/challenge2022/task-language-based-audio-retrieval}. Firstly, we introduce a simple, scalable architecture which ties both the audio and text encoder together. Secondly, we show that using this architecture along with contrastive loss allows the model to significantly beat the performance of the baseline model. Finally, in addition to having an extremely low training memory requirement, we are able to use pretrained models as it is without needing to finetune them. We test our methods and show that using a combination of our methods beats the baseline scores significantly.

\end{abstract}
%
%
\section{Introduction}
\label{sec:intro}

The Language-Based Audio Retrieval task is a new form of cross modal learning \cite{Xie_2022_audio_retrieval} which aims to rank a list of audio clips according to their relevance given a query caption. These query captions \cite{drossos_clotho_2019} are descriptive natural language sentences annotated by humans, and they describe the acoustic events happening in both foreground and background of the audio clip. Being able to model and interpret the relationship between audio clips and a text sequence is helpful towards many applications. Language-Based Audio Retrieval can be applied to many practical applications in real life, such as acoustic monitoring and human-computer interaction \cite{Xie_2022_audio_retrieval}.

The Language-Based Audio Retrieval task is formulated in this way. The audio clips and the query caption is passed to an audio encoder and text encoder respectively. From the two encoders, output audio embeddings corresponding to audio clips and a output text embedding corresponding to the query caption are obtained. To determine the relevance between the audio clip and query caption, a similarity measure is used to calculate similarity scores between the output text embedding of the query caption and the output audio embedding of each audio clip. Using these similarity scores, we can calculate the top 10 average precision, and the top 1, top 5, and top 10 recall scores.

The baseline system proposed for Language-Based Audio Retrieval in DCASE2022 uses a dual encoder structure with two disjoint pathways to produce the output audio and text embeddings. The CRNN model \cite{crnn_gru_baseline, text-to-audio-grounding} was used as the audio encoder while the pretrained word2vec \cite{word2vec} model was used as the text encoder. The model is trained using the Triplet ranking Loss \cite{triplet_ranking} to maximize the distance between the anchor sample and the negative sample, while minimizing the distance between the anchor sample and the positive sample. During inference, the output audio and text embeddings are extracted from the encoders. Then, the dot product is used as the similarity measure to determine the relevance of the audio clips to the query caption. We argue that training to maximize the similarity between the output embeddings of two disjoint and separate encoders for audio and text in Language-Based Audio Retrieval is non-trivial. We find that we can increase efficiency and performance by tying the audio and text encoder together and sharing their parameters. In addition to having a tied model to produce output embeddings, we find that using contrastive loss is instrumental in getting the model to converge and perform well. Finally, we show compare the computational footprint of our methods and show the efficiency of our method.

Our contributions are as follows:

\begin{enumerate}
    \item We introduce Converging Tied Layers and show that using Converging Tied Layers for Language-Based Audio Retrieval is an efficient and straightforward method.
    \item We examine the importance of using contrastive loss and observe that contrastive loss is crucial for using transformers effectively.
    \item We demonstrate that using Converging Tied Layers and contrastive loss outperforms the baseline method by a significant factor,
\end{enumerate}

\section{Related Work} \label{sec:related_work }

\subsection{Datasets} 
\label{sec:datasets}
The Clotho Dataset v2.1 consists of 6974 15 to 30 seconds long audio samples. Each audio clip has 5 corresponding 8 to 20 words long human-annotated captions that describe the acoustic events happening in the audio. During training, the ground truth captions for that audio are used as the positive samples and the ground truth captions for the other audios are used as the negative samples. During evaluation, all of the audio clips are passed to the model to rank each audio's similarity to the query caption. Another dataset is the Audio Grounding dataset has also been used by \cite{text-to-audio-grounding} for Audio and Caption Retrieval. Though there are many other audio captioning datasets where Language-Based Audio Retrieval can be applied to, to our knowledge active work is ongoing only on the Clotho dataset and Audio Grounding dataset. 

In this work, we focus only on the Clotho Dataset v2.1 as proposed in the DCASE2022 challenge\footnote{https://dcase.community/challenge2022/task-language-based-audio-retrieval}, henceforth referred to as the Clotho Dataset.

\subsection{Model Architectures}
\label{model_arch_objectives}

Prior work so far uses disjoint audio and text encoders to produce a vector representation of the inputs. The baseline model \cite{crnn_gru_baseline} presented in DCASE 2022 uses 2 disjoint audio and text models to encode the audio clip and text from the Clotho Dataset v2.1. The input audio is encoded by a Convolutional Recurrent Neural Network (CRNN) \cite{crnn_gru_baseline} and is trained from scratch. For the input text sequence, a pretrained word2vec \cite{word2vec} model\footnote{https://code.google.com/archive/p/word2vec/} already trained on the Google News dataset \cite{googlenews} is used to encode the text sequence to obtain a text vector representation. The pretrained word2vec is not finetuned. \cite{crnn_gru_baseline} also uses a similar approach for the Audio Grounding Dataset.

\subsection{Contrastive Loss}

During the advent of large scale pretrained language models\cite{devlin2019bert, roberta}, many authors focused on different predictive objectives such as masked language modelling \cite{devlin2019bert,roberta} for pretraining. Over time, the focus shifted to a different form of using contrastive loss to learn more informative multimodal embedding spaces. There has been several variations of contrastive loss \cite{clip,audioclip,contrastive_medical}. In this work, we use the contrastive objective from CLIP \cite{clip, audioclip}. 

The CLIP contrastive objective extracts feature representations of the different input modalities from the model and projects these representation to a contrastive embedding space via a linear projection. The projected representations are then normalized. The cosine similarities between every possible pairwise representations of different modalities in the same batch are calculated to obtain logits for both text and audio. Finally, the contrastive loss is the average of the two cross entropy loss applied to the text and audio logits with their labels being their respective index in the batch.

\begin{figure}[ht!]
    \centering
    \includegraphics[scale=0.40]{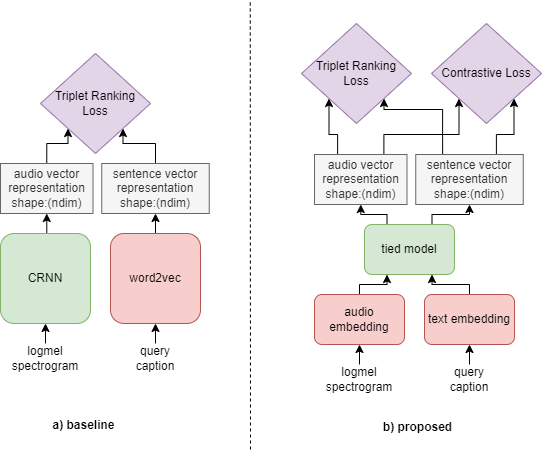}
    \caption{\textbf{a)} Baseline system. A CRNN is trained for the audio encoder while a word2vec model pretrained on Google News is used without any finetuning. \textbf{b)} Proposed system. Both audio embeddings and text embeddings are used with frozen weights without any finetuning. We use CNN10, CNN14 for the audio embeddings and BERT, RoBERTa for the text embeddings. Both embeddings are passed to the tied model which is trained on both Triplet Ranking Loss and Contrastive Loss. Shaded red boxes in the figure refers to models with frozen parameters (not finetuned) while green boxes refers to layers/models with trainable parameters.}
    \label{fig:training}
\end{figure}

\begin{figure}[ht!]
    \centering
    \includegraphics[scale=0.45]{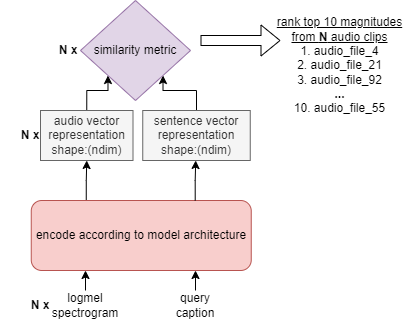}
    \caption{Evaluation process: The logmel spectrogram of each audio clip in the evaluation set is encoded to obtain its corresponding vector representation. The query caption is likewise encoded to obtain the sentence vector representation. The similarity metric is then applied between each audio representation and the sentence representation to obtain a list of similarity values. These values are then ranked to obtain the relevance of each audio clip to the query caption.}
    \label{fig:inference}
\end{figure}

\section{Proposed Method} \label{sec:proposed_method}

Our proposed model consists of two main parts. The first component refers to the use of pretrained audio and text encoders as audio and text embeddings. There is no finetuning of these embeddings and the weights of these encoders are frozen. The second component is the Converging Tied layers. These layers are shared between the audio and text input. Unlike prior work \cite{text-to-audio-grounding}, where the output embeddings are extracted via two disjoint and seperate models, we use the same layers to produce both audio and text embeddings. We visualize this in Figure \ref{fig:training} and \ref{fig:inference}.

\subsection{Pretrained Embeddings}
There is a plethora of pretrained models available publicly. These pretrained models are often used for transfer learning to another related domain \cite{koh2021automated}, hence there is a need for finetuning. In our case, we find that it is sufficient to simply use these pretrained models as it is without finetuning. Therefore, there is a very low computational footprint from these embeddings. However, we also performed some experiments where we finetuned on these pretrained embeddings and we found that doing so yields a minimal performance boost.

\begin{equation}
  \begin{array}{l}
    Emb_{A} = \text{pool}_{mean}(Encoder\textsubscript{A}(x_{A})) \\
    Emb_{T} = Encoder\textsubscript{T}(x_{T}) \\
  \end{array}
 \label{eqn:tied_layers1}
\end{equation}

We use the CNN10 and CNN14 models already pretrained on audio tagging as the audio encoder to produce audio embeddings, $Emb_{A}$. For the text embeddings, we use BERT and RoBERTa as the text encoder to produce text embeddings, $Emb_{T}$. Unless otherwise stated, these pretrained embeddings are frozen and not finetuned, thereby minimizing the training time.

\subsection{Converging Tied layers}

The Converging Tied Layers take in both the audio embedding, $Emb_{A}$, and text embedding, $Emb_{T}$, and project these embeddings to a common subspace. We first pass both $Emb_{A}$ and $Emb_{T}$ through a linear layer for each modality to project these embeddings to the same dimensionality. The projected audio and text embeddings, $R_{A^\prime}$ and $R_{T^\prime}$, are then passed through several shared layer to obtain the final vector representations of the audio and text inputs, $R_A$ and $R_T$. While we defaulted to transformer encoder layers due to its ability to encode contextual information, we also experimented with simple feedforward layers. 

\begin{equation}
  \begin{array}{l}
    R_{A^\prime} = \text{FFN}_A(Emb_{A}) \\
    R_{T^\prime} = \text{FFN}_T(Emb_{T}) \\
    
    R_A = \text{pool}_{mean}(\text{Transformer}_{tied}(R_{A^\prime})) \\
    R_T = \text{pool}_{mean}(\text{Transformer}_{tied}(R_{T^\prime}))

  \end{array}
 \label{eqn:tied_layers2}
\end{equation}

These tied layers share parameters across text and audio inputs and produces both the audio and text vector representations. We hypothesize that using a shared embedding subspace allows the model to perform better on the ranking task, as opposed to having two disjoint encoders with two disjoint embedding subspace.

\subsubsection{Contrastive Loss} \label{sec:contrastive_loss}
In addition to the Triplet Ranking Loss used, we also use a supplementary Contrastive Loss jointly train the model. We find that using the Contrastive loss is instrumental in helping the model converge. We use the same contrastive loss as that in CLIP \cite{clip,audioclip, contrastive_medical}. 

\begin{equation}
\begin{array}{l}
    L = L_{Ranking} + L_{contrastive} \\
 \end{array}
 \label{eqn:loss_combination}
\end{equation}

The model is trained to minimize both the triplet ranking loss, $L_{Ranking}$, from positive and negative examples in the minibatch, and the contrastive loss, $L_{contrastive}$, from the predicting the correct pair in the batch \cite{contrastive_medical}.

\section{Experimental Details} \label{sec:experimental_details}
\subsection{Data}
We use Clotho dataset v2.1 for all our experiments as mentioned in Section \ref{sec:datasets}. We extract log mel spectrogram with 64 Mel-bands, sampling rate 44100, and hop length of 441. This is identical to the settings of the previous baseline.

\subsection{Training and Evaluation} \label{sec:training}

Our training hyperparameters and settings are as follows. We use a batch size of 32 with no gradient accumulation steps for 150 epochs with early stopping based on the validation performance. We use a learning rate of $1 \times 10^{-3}$ without any weight decay along with a learning rate scheduler which reduces the learning rate with a factor of 0.1 when the performance plateaus for 5 epochs. For the audio embeddings, we initialized the weights of the pretrained CNN10 and CNN14 model\footnote{Publicly available at https://github.com\/qiuqiangkong\/audioset\_tagging\_cnn}. For the text embeddings, we used the pretrained BERT and RoBERTa model provided by Hugging Face\footnote{huggingface.co/docs/transformers}. Unless explicitly stated, these pretrained embeddings are frozen in our experiments and the weights are not updated.

For inference, the dot product between the vector representations of the audio clips in the evaluation set and the vector representation of each query caption in the evaluation set is used as the similarity measure to determine the relevance of the audio clip to the query caption. The metrics used to gauge performance are mean average precision at 10, and top 1, top 5 and top 10 recall.

\section{Experimental Results and Analysis} \label{sec:experimental_results_analysis}
We provide a summary of our best models and methods in Table \ref{tab:results_comparison_best}. In the following sections, we will analyse and provide some ablation studies of our methods along with more comphrensive results. As mentioned in Section \ref{sec:training}, the pretrained weights of the audio and text embeddings are frozen unless otherwise stated.

\begin{table}[!ht]
\centering
\resizebox{\columnwidth}{!}{\begin{tabular}{@{}lllllll@{}}
\toprule
Encoder\textsubscript{A} & Encoder\textsubscript{T} & Tied Model & R\textsubscript{1} & R\textsubscript{5} & R\textsubscript{10} & mAP\textsubscript{10}\\ \midrule
CRNN &	word2vec & - &0.03 &	0.11&	0.19&	0.07\\
\midrule
CNN10 &	RoBERTa\textsubscript{base}& 4L 96dim Transformer & 0.10&	0.29&	0.41&	0.18\\
\textit{CNN10} &	\textit{{RoBERTa\textsubscript{base}}}& \textit{2L 192dim Transformer} &\textit{0.11}&	\textit{0.30}&
\textit{0.42}&	\textit{0.19}\\
\textbf{\textit{CNN10}} &	\textit{\textbf{RoBERTa\textsubscript{base}}} & \textit{\textbf{4L 96dim Transformer}} & \textit{\textbf{0.11}}&	\textit{\textbf{0.32}}&	\textit{\textbf{0.45}}&	\textit{\textbf{0.2}}\\
\bottomrule
\end{tabular}
}
\caption{Comparison of our best 3 models against the baseline model (1st row). Bold (4th row) indicates our best performing model. Italics (3rd and 4th row) indicate that the model is fully trainable and no weights are frozen.}
\label{tab:results_comparison_best}
\end{table}

\subsection{Importance of the Contrastive loss}

\begin{table}[!ht]
\centering
\resizebox{\columnwidth}{!}{\begin{tabular}{@{}lllllll@{}}
\toprule
Encoder\textsubscript{A} & Encoder\textsubscript{T} & Tied Model & R\textsubscript{1} & R\textsubscript{5} & R\textsubscript{10} & mAP\textsubscript{10}\\ \midrule
CRNN &	word2vec & - &0.03 &	0.11&	0.19&	0.07\\
\midrule
CNN10 &	RoBERTa\textsubscript{base} & 2L 192dim Linear & 0.00&	0.00&	0.01 & 0\\
CNN10 &	RoBERTa\textsubscript{large} & 2L 192dim Linear & 0.00&	0.00&	0.01 & 0\\
CNN10 &	BERT\textsubscript{base} & 2L 192dim Linear & 0.01&	0.05&	0.10 & 0.03\\
CNN10 &	BERT\textsubscript{large} & 2L 192dim Linear & 0.00& 0.00&	0.01 & 0\\

\bottomrule
\end{tabular}}
\caption{Models trained without contrastive loss. Without contrastive loss, the model fails to perform well.}
\label{tab:results_comparison_contrastive}
\end{table}

As mentioned in Section \ref{sec:contrastive_loss}, we use a supplementary contrastive loss in addition to the Triplet Ranking Loss. We find that without the contrastive loss, the model is unable to converge and performs very badly. Our results are shown in Table \ref{tab:results_comparison_contrastive}. For all other experiments, we defaulted to using contrastive loss as supplementary objective.

\subsection{Impact of using trainable or frozen pretrained embeddings}

\begin{table}[!ht]
\centering
\resizebox{\columnwidth}{!}{\begin{tabular}{@{}lllllll@{}}
\toprule
Encoder\textsubscript{A} & Encoder\textsubscript{T} & Tied Model & R\textsubscript{1} & R\textsubscript{5} & R\textsubscript{10} & mAP\textsubscript{10}\\ \midrule
CRNN &	word2vec & - &0.03 &	0.11&	0.19&	0.07\\
\midrule
CNN10 &	BERT\textsubscript{large} & 2L 192dim Transformer & 0.06&	0.20&	0.31 &	0.12\\
CNN10 &	BERT\textsubscript{base} & 2L 192dim Transformer & 0.07&	0.23&	0.34 &	0.14\\
\midrule
CNN10 &	RoBERTa\textsubscript{large} & 2L 192dim Transformer & 0.08&	0.24&	0.37 &	0.15\\
CNN14 &	RoBERTa\textsubscript{base} & 2L 192dim Transformer & 0.09&	0.26&	0.37 &	0.16\\
CNN10 &	RoBERTa\textsubscript{base} & 2L 192dim Transformer & 0.10&	0.28&	0.40 &	0.18\\
\midrule
\textit{CNN10} &	\textit{RoBERTa\textsubscript{base}}& \textit{2L 192dim Transformer} &\textit{0.11}&\textit{0.30}&	\textit{0.42}&	\textit{0.19}\\
\bottomrule
\end{tabular}}
\caption{Comparison of the choice of pretrained embeddings for the audio and text embeddings. Italics (last row) indicate that the model is fully trainable and no weights are frozen.  }
\label{tab:results_comparison_trainable}
\end{table}


We compare the effectiveness of using pretrained embedding. Results are shown in Table \ref{tab:results_comparison_trainable}. We experimented with using CNN10/CNN14 \cite{kong2020panns} as the audio embeddings and Bert/RoBERTa as the text embeddings.  Using either trainable or frozen pretrained embeddings with the Converging Tied Layers both surpass the baseline performance significantly. We also note that trainable pretrained embeddings do perform marginally better than their frozen counterparts at the cost of more computational power and memory.  

RoBERTa consistently performs significantly better than BERT, even though their model sizes are similar. This is expected as RoBERTa \cite{roberta} outperforms BERT on many Natural Language Processing benchmarks such as GLUE \cite{glue_benchmark}, RACE \cite{race_benchmark}, SQuaD \cite{squad_benchmark}. Therefore, RoBERTa is regarded as a better and more robust model. This indicates that initialization of the embeddings is important and any information stored in pretrained embeddings helps the model perform better for audio retrieval. We also observe that smaller variants of the pretrained audio and embeddings perform significantly better than their larger variants. For instance, BERT\textsubscript{base} and RoBERTa\textsubscript{base} perform better than BERT\textsubscript{large} and RoBERTa\textsubscript{large} with around 0.03 difference in mAP\textsubscript{10}.

\subsection{Tied Transformers layers vs Tied Linear layers}

\begin{table}[!ht]
\centering
\resizebox{\columnwidth}{!}{
\begin{tabular}{@{}lllllll@{}}
\toprule
Encoder\textsubscript{A} & Encoder\textsubscript{T} & Tied Model & R\textsubscript{1} & R\textsubscript{5} & R\textsubscript{10} & mAP\textsubscript{10}\\ \midrule
CRNN &	word2vec & - &0.03 &	0.11&	0.19&	0.07\\
\midrule
CNN14 &	RoBERTa\textsubscript{base} & 2L 192dim Linear & 0.03&	0.15&	0.24 & 0.08\\
CNN14 &	RoBERTa\textsubscript{base} & 2L 300dim Linear & 0.03&	0.14&	0.24 & 0.08\\
CNN14 &	RoBERTa\textsubscript{base} & 2L 192dim Linear & 0.03&	0.14&	0.23 & 0.08\\
\midrule
CNN14 &	RoBERTa\textsubscript{base} & 2L 192dim Transformer & 0.09&	0.26&	0.37 &	0.16\\
\bottomrule
\end{tabular}}
\caption{Comparison of the choice of Converging Tied Layers. Converging Tied Linear layers consistently gets outperformed by Converging Tied Transformers layers.}
\label{tab:results_comparison_tied}
\end{table}

We explore the choice of the type of layer to use for Converging Tied Layers. Results are shown in Table \ref{tab:results_comparison_tied}. Transformers are known for being able to encode contextual information \cite{wav2vec2} while linear layers provide a transformation between features. In our experiments, Converging Tied Linear layers consistently gets outperformed by Converging Tied Transformers layers. This confirms our hypothesis that using the transformer encoder layers as the choice for Converging Tied layers allows the model to better interpolate contextual information from both the audio embedding and text embedding to a common subspace.

\section{Conclusion} \label{sec:conclusion}
This work introduces the use of Converging Tied layers and the importance of contrastive loss for Language Basd Audio Retrieval. We show that Converging Tied layers is a straightforward and efficient method that allows for minimal training. We examined and analysed several design choices such as the choice for converging tied layers and also the preference for smaller embeddings. With our methods, we surpass the baseline model significantly on all metrics.

\bibliographystyle{IEEEbib}
\bibliography{refs}

\end{document}